\documentclass[journal]{IEEEtran}

\usepackage{amsmath}
\usepackage{algorithmic}
\usepackage{array}
\usepackage[usenames,dvipsnames]{pstricks}
 \usepackage{epsfig}
 \usepackage{pst-grad} 
 \usepackage{pst-plot} 
\usepackage{dsfont,amsfonts,latexsym,amsmath,amssymb}
\usepackage{amsfonts}
\usepackage{graphicx}
\usepackage{amsmath}
\usepackage{gensymb}
\usepackage{tikz}
\newcommand\scalemath[2]{\scalebox{#1}{\mbox{\ensuremath{\displaystyle #2}}}}

\def\EMDD{E\left(\theta_x,\theta_y\right)=\frac{R}{J}\begin{bmatrix}
-\sin\phi_1\cos\theta_x\cos\theta_y & -\sin\phi_2\cos\theta_x\cos\theta_y\\
\cos\phi_1\cos\theta_y+\sin\phi_1\sin\theta_x\sin\theta_y & \cos\phi_2\cos\theta_y+\sin\phi_2\sin\theta_x\sin\theta_y
\end{bmatrix}
}
\def\GMDD{G\left(\theta_x,\theta_y\right)=\frac{R}{J}\begin{bmatrix}
-\sin\phi_3\cos\theta_x\cos\theta_y\\
\cos\phi_3\cos\theta_y+\sin\phi_3\sin\theta_x\sin\theta_y
\end{bmatrix}
}

\newtheorem{the:def}{Definition}	 	
\newtheorem{the:lem}[the:def]{Lemma}	
\newtheorem{the:the}[the:def]{Theorem}	

\begin{document}

\title{Flatness-based control of a two-degree-of-freedom platform with  pneumatic artificial muscles}

\author{David Bou Saba,
        Paolo Massioni,
        Eric Bideaux,
        and Xavier Brun
\thanks{The authors are with Laboratoire Amp\`ere, UMR CNRS 5005, INSA de Lyon, Universit\'e de Lyon, 69621 Villeurbanne CEDEX, France. {\tt\small  \{david.bou-saba, paolo.massioni, eric.bideaux, xavier.brun\}@insa-lyon.fr}. Corresponding author: Paolo Massioni, tel: +33(0)472436035, fax: +33(0)472438530.}
}


\maketitle

\begin{abstract}
Pneumatic artificial muscles are a quite interesting type of actuators which have a very high power-to-weight and power-to-volume ratio. However, their efficient use requires very accurate control methods which can take into account their complex dynamic, which is highly nonlinear. This paper consider a model of two-degree-of-freedom platform whose attitude is determined by three pneumatic muscles controlled by servovalves, which mimics a simplified version of a Stewart platform. For this testbed, a model-based control approach is proposed, based on accurate first principle modeling of the muscles and the platform and on a static model for the servovalve. 
The employed control method  is the so-called flatness-based control introduced by Fliess. The paper first recalls the basics of this control technique and then it shows how it can be applied to the proposed experimental platform; being flatness-based control an open-loop kind of control, a proportional-integral controller is added on top of it in order to add robustness with respect to modelling errors and external perturbations. 
At the end of the paper, the effectiveness of the proposed approach is shown by means of experimental results. A clear improvement of the tracking performance is visible compared to a simple proportional-integral controller.

\end{abstract}



\section{Introduction}

Pneumatic artificial muscles (PAMs) are a quite efficient type of actuators which feature high power-to-volume ratio, high pulling efforts at a relatively low price  \cite{daerden2002pneumatic}. This makes their use quite interesting in  many engineering and robotic applications, even if their control is problematic due the non-linearity in their  dynamic model  as well as from the hysteresis phenomena which they feature. 

Pneumatic artificial muscles produce a contraction effort when inflated, which is a nonlinear function of both the internal pressure and the relative contraction of its length. Many theoretical models of  PAMs can be found in the literature \cite{daerden2002pneumatic, chou1996measurement, tondu2000modeling}, and this paper will refer to the results of experimental tests \cite{fluidpower} that average out the hysteresis phenomena  and therefore can model the behaviour very accurately.  The subject of this paper is a study of a two-degree-of-freedom platform, actuated by three pneumatic muscles. The objective is the synthesis of a model-based control law allowing the tracking of a reference trajectory for a wide operating range of the muscles. The platform is constrained to a limited operating domain due to mechanical constraints and to the fact that the muscles generate only pulling efforts. Furthermore, the system can be considered as overactuated (three actuators moving two degrees of freedom), which requires a control allocation strategy.
 
The control of PAMs has been approached with several methods, which try to cope with the strong nonlinearities of its dynamics. The approaches found in the literature are mainly inherently nonlinear control methods \cite{ba2016integrated,zhu2008adaptive}; sliding mode controllers are one of the most common choices \cite{aschemann2008sliding,cai2000sliding,shen2010nonlinear}, also sometimes combined with adaptive or  neural controllers \cite{shi2013hybrid,robinson2016nonlinear}, or backstepping \cite{abd2016design}. Sliding mode controllers in fact provide enough robustness with respect to the dynamical model which is considered as uncertain.

In this work,  a flatness-based control \cite{fliess1995flatness} is proposed, which exploits a model of all the elements involved and which also solves the over-actuation problem at the same time. The robustness with respect to model errors is provided by coupling the flatness-based controller with a proportional-integral (PI) controller feeding back the error with respect to the reference trajectory.

The paper is structured as follows.
Section~\ref{sec:nota} introduces  the notation used throughout the paper. 
Section~\ref{sec:mode} describes the model of the platform and of all its elements, including the pneumatic artificial  muscles. 
Section~\ref{sec:anal} shows that a proper choice of measurements makes the platform a flat system, for which a flatness-based law is proposed. 
Section~\ref{sec:over} concerns the problem of overactuation and how it is solved. At last,
Section~\ref{sec:expe} proposes some experimental results whereas
Section~\ref{sec:conc} draws the conclusions of the article.

\section{Notation and definitions}
\label{sec:nota}
Let \(\mathbb{R}\) be the set of real number, and \(\mathbb{N}\) the set of the strictly positive integers. For a matrix \(A\), \(A^\top\) denotes the transpose. Given two functions \(f(x), g(x) \in \mathbb{R}^n \), with \(x \in \mathbb{R}^n\), let the Lie derivative of \(f\) along \(g\) be defined as \(L_g f(x) =\frac{ \partial f(x)}{\partial x} \cdot g(x)\). For \(\xi \in \mathbb{N}\), let \(L_g^{\xi} f(x) =\frac{ \partial L_g^{\xi-1} f(x)}{\partial x} \cdot g(x)\), with \(L_g^0 f(x) = f(x)\). For all signals \(x\) depending from the time \(t\), let \(x^{(\xi)}\) indicate its \(\xi\)-th time derivative, i.e. \(x^{(1)} = \frac{dx}{dt}= \dot{x}\),   \(x^{(2)} = \frac{d^2x}{dt^2}= \ddot{x}\), etc.

All the symbols concerning the pneumatic muscle platform are defined in Table~\ref{tab:syb}.

\begin{table}[h]\centering
\begin{tabular}{|ll|}
\hline
{$P_0$}&{Atmospheric pressure}\\
{$\theta_0$}&{Weave angle of the muscle at rest}\\
{$D_0$}&{Diameter of the muscle at rest}\\
{$l_0$}&{Length of the muscles at rest}\\
{$\alpha$}&{Experimentally determined power coefficient}\\
{$K$}&{Experimentally determined coefficient}\\
{$\varepsilon_a$}&{Experimentally determined coefficient}\\
{$\varepsilon_b$}&{Experimentally determined coefficient}\\
{$k$}&{Polytropic index of air }\\
{$r$}&{Perfect gas constant}\\
{$T$}&{Air temperature}\\
{$R$}&{Muscle application point distance from center (constant)}\\
{$J$}&{Momentum of inertia about an horizontal axis (constant)}\\
{$\phi_1=-90^\circ$}&{Angular position of the \(1\)st  muscle (constant)}\\
{$\phi_2=30^\circ$}&{Angular position of the \(2\)nd  muscle (constant)}\\
{$\phi_3=150^\circ$}&{Angular position of the \(3\)rd  muscle (constant)}\\
{$\theta_x$}&{Angular position of the platform around $x$ axis}\\
{$\theta_y$}&{Angular position of the platform around $y$ axis}\\
{$P_i$}&{Absolute pressure inside the \(i\)-th muscle}\\
{$v_i$}&{Voltage applied to the \(i\)-th servovalve}\\
{$V_i$}&{Volume of the \(i\)-th muscle}\\
{$q_i$}&{Mass flow into the \(i\)-th muscle}\\
{$\varepsilon_i$}&{Contraction of the \(i\)-th muscle}\\
{$\varepsilon_0$}&{Initial contraction of the muscle}\\
{$F_i$}&{Force applied by the \(i\)-th the muscle}\\
{$\Gamma$}&{Perturbation torques} \\\hline
\end{tabular}
\caption{Symbol definitions.}
\label{tab:syb}
\end{table}

\section{The pneumatic platform}
\label{sec:mode}

\subsection{Description}
\begin{figure}[h]
\centering
\includegraphics[width=\columnwidth]{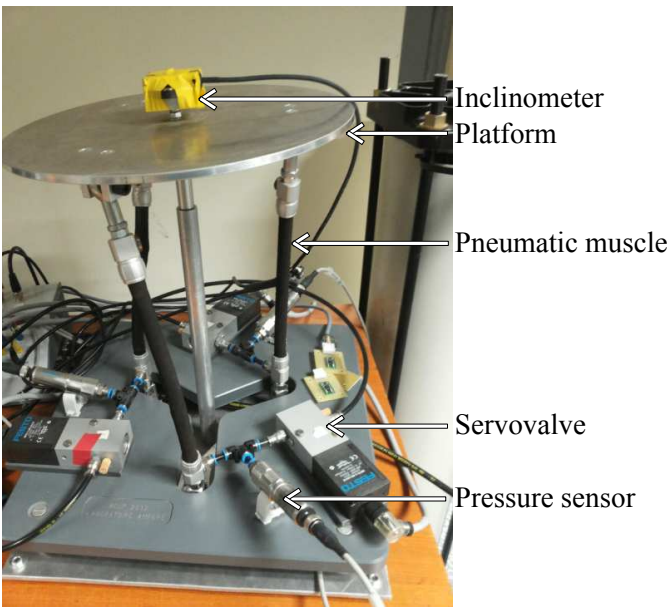} 
\caption{The experimental platform.} 
\label{fig:platform}
\end{figure}

\begin{figure}[h]
\centering
\scalebox{0.60} 
{
\begin{pspicture}(0,-3.0789063)(26.341875,3.0589063)
\definecolor{color872b}{rgb}{0.4,0.4,0.4}
\psline[linewidth=0.04cm,arrowsize=0.05291667cm 2.0,arrowlength=1.4,arrowinset=0.4]{->}(10.1,-0.26109374)(13.3,-0.26109374)
\pscircle[linewidth=0.04,dimen=outer](10.1,-0.26109374){2.3}
\psline[linewidth=0.04cm,arrowsize=0.05291667cm 2.0,arrowlength=1.4,arrowinset=0.4]{->}(10.1,-0.26109374)(10.1,3.0389063)
\rput{22.5}(0.1594943,-1.2886641){\psellipse[linewidth=0.04,dimen=outer](3.3190234,-0.24341609)(3.0731425,1.2538031)}
\psline[linewidth=0.04cm,arrowsize=0.05291667cm 2.0,arrowlength=1.4,arrowinset=0.4]{->}(3.3,-0.26109374)(6.9,-0.26109374)
\psline[linewidth=0.04cm,arrowsize=0.05291667cm 2.0,arrowlength=1.4,arrowinset=0.4]{->}(3.3,-0.26109374)(5.6,2.0389063)
\pscircle[linewidth=0.04,dimen=outer](10.1,-0.26109374){0.2}
\pscircle[linewidth=0.04,dimen=outer](3.3,-0.26109374){0.2}
\psline[linewidth=0.04cm](2.9,-0.46109375)(3.7,-0.46109375)
\psline[linewidth=0.04cm](2.9,-0.6610938)(3.1,-0.46109375)
\psline[linewidth=0.04cm](3.1,-0.6610938)(3.3,-0.46109375)
\psline[linewidth=0.04cm](3.3,-0.6610938)(3.5,-0.46109375)
\psline[linewidth=0.04cm](3.5,-0.6610938)(3.7,-0.46109375)
\psline[linewidth=0.04cm](9.7,-0.6610938)(9.9,-0.46109375)
\psline[linewidth=0.04cm](9.9,-0.6610938)(10.1,-0.46109375)
\psline[linewidth=0.04cm](10.1,-0.6610938)(10.3,-0.46109375)
\psline[linewidth=0.04cm](10.3,-0.6610938)(10.5,-0.46109375)
\psline[linewidth=0.04cm](9.7,-0.46109375)(10.5,-0.46109375)
\psline[linewidth=0.04cm](3.1,-0.26109374)(3.0,-0.46109375)
\psline[linewidth=0.04cm](3.5,-0.26109374)(3.6,-0.46109375)
\psline[linewidth=0.04cm](9.9,-0.26109374)(9.8,-0.46109375)
\psline[linewidth=0.04cm](10.3,-0.26109374)(10.4,-0.46109375)
\psarc[linewidth=0.04,arrowsize=0.05291667cm 2.0,arrowlength=1.4,arrowinset=0.4]{<-}(5.6,2.0389063){0.3}{0.0}{180.0}
\psarc[linewidth=0.04](3.35,-0.31109375){0.05}{0.0}{180.0}
\psline[linewidth=0.04cm,arrowsize=0.05291667cm 2.0,arrowlength=1.4,arrowinset=0.4]{->}(3.3,-0.26109374)(3.3,-2.1610937)
\rput{45.5}(1.7621037,-4.7240853){\psarc[linewidth=0.04,arrowsize=0.05291667cm 2.0,arrowlength=1.4,arrowinset=0.4]{->}(6.5138845,-0.2609726){0.29018185}{341.86905}{180.0}}
\psline[linewidth=0.08cm,linestyle=dashed,dash=0.16cm 0.16cm,arrowsize=0.05291667cm 2.0,arrowlength=1.4,arrowinset=0.4]{->}(2.2,-1.5610938)(2.2,-2.6610937)
\psline[linewidth=0.08cm,linestyle=dashed,dash=0.16cm 0.16cm,arrowsize=0.05291667cm 2.0,arrowlength=1.4,arrowinset=0.4]{->}(5.4,0.43890625)(5.4,-0.86109376)
\psline[linewidth=0.08cm,linestyle=dashed,dash=0.16cm 0.16cm,arrowsize=0.05291667cm 2.0,arrowlength=1.4,arrowinset=0.4]{->}(2.5,0.43890625)(2.5,-0.86109376)
\psline[linewidth=0.08cm,arrowsize=0.05291667cm 2.0,arrowlength=1.4,arrowinset=0.4]{->}(2.2,-1.8610938)(2.2,-2.6610937)
\psline[linewidth=0.08cm,arrowsize=0.05291667cm 2.0,arrowlength=1.4,arrowinset=0.4]{->}(5.4,-0.46109375)(5.4,-0.86109376)
\psellipse[linewidth=0.04,dimen=outer,fillstyle=solid,fillcolor=color872b](2.5,0.43890625)(0.1,0.1)
\psellipse[linewidth=0.04,dimen=outer,fillstyle=solid,fillcolor=color872b](5.4,0.43890625)(0.1,0.1)
\psellipse[linewidth=0.04,dimen=outer,fillstyle=solid,fillcolor=color872b](2.2,-1.3610938)(0.1,0.1)
\usefont{T1}{ptm}{m}{n}
\rput(5.541406,2.6489062){$\theta_y$}
\usefont{T1}{ptm}{m}{n}
\rput(6.5314064,0.24890625){$\theta_x$}
\usefont{T1}{ptm}{m}{n}
\rput(5.6514063,1.7489063){$y$}
\usefont{T1}{ptm}{m}{n}
\rput(6.8414063,-0.45109376){$x$}
\usefont{T1}{ptm}{m}{n}
\rput(2.1414063,-2.8510938){$F_1$}
\usefont{T1}{ptm}{m}{n}
\rput(5.3414063,-1.0510937){$F_2$}
\usefont{T1}{ptm}{m}{n}
\rput(2.1414063,-0.75109375){$F_3$}
\usefont{T1}{ptm}{m}{n}
\rput(1.8014063,-1.3510938){$M_1$}
\usefont{T1}{ptm}{m}{n}
\rput(4.9014063,0.44890624){$M_2$}
\usefont{T1}{ptm}{m}{n}
\rput(3.0014062,0.44890624){$M_3$}
\psellipse[linewidth=0.04,dimen=outer,fillstyle=solid,fillcolor=color872b](10.1,-1.8610938)(0.1,0.1)
\psellipse[linewidth=0.04,dimen=outer,fillstyle=solid,fillcolor=color872b](11.7,0.5389063)(0.1,0.1)
\psellipse[linewidth=0.04,dimen=outer,fillstyle=solid,fillcolor=color872b](8.5,0.5389063)(0.1,0.1)
\psline[linewidth=0.02cm](10.1,-0.26109374)(11.7,0.5389063)
\psline[linewidth=0.02cm](10.1,-0.26109374)(8.5,0.5389063)
\psline[linewidth=0.02cm](10.1,-0.26109374)(10.1,-1.8610938)
\usefont{T1}{ptm}{m}{n}
\rput(10.101406,-2.1510937){$M_1$}
\usefont{T1}{ptm}{m}{n}
\rput(11.501407,0.8489063){$M_2$}
\usefont{T1}{ptm}{m}{n}
\rput(8.7014065,0.8489063){$M_3$}
\usefont{T1}{ptm}{m}{n}
\rput(10.4514065,2.5489063){$y$}
\usefont{T1}{ptm}{m}{n}
\rput(13.041407,0.04890625){$x$}
\psarc[linewidth=0.02,arrowsize=0.05291667cm 2.0,arrowlength=1.4,arrowinset=0.4]{->}(10.1,-0.26109374){1.6}{0.0}{27.0}
\psarc[linewidth=0.02,arrowsize=0.05291667cm 2.0,arrowlength=1.4,arrowinset=0.4]{->}(10.1,-0.26109374){1.4}{0.0}{153.0}
\psarc[linewidth=0.02,arrowsize=0.05291667cm 2.0,arrowlength=1.4,arrowinset=0.4]{<-}(10.1,-0.26109374){1.2}{270.0}{0.0}
\usefont{T1}{ptm}{m}{n}
\rput(10.311406,-1.1510937){$\phi_1$}
\usefont{T1}{ptm}{m}{n}
\rput(11.9114065,0.04890625){$\phi_2$}
\usefont{T1}{ptm}{m}{n}
\rput(9.211407,0.44890624){$\phi_3$}
\usefont{T1}{ptm}{m}{n}
\rput(3.6414063,-2.0510938){$z$}
\end{pspicture} 
}
\caption{Axonometric view and view from the top of the top plate, with definition of the axes $x$, $y$, $z$ and the rotation angles $\theta_x$ and $\theta_y$. $M_1$, $M_2$ and $M_3$ are the attachment points of the three pneumatic artificial muscles.} 
\label{fig:plate}
\end{figure}
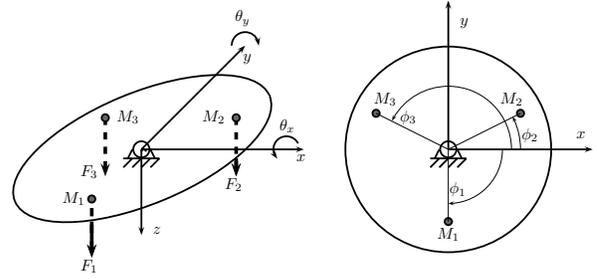

The pneumatic platform  studied in this paper is represented in Fig.~\ref{fig:platform} and Fig.~\ref{fig:plate}. It consists of a metal plate fixed to a spherical hinge on top of a vertical beam;  three pneumatic muscles controlled by servovalves are attached to the plate at equally spaced points. Due to the muscles, and for simplicity, it can be considered that the platform has only two degrees of freedom, i.e. the two rotational angles ($\theta_x$ and $\theta_y$) with respect to horizontal axes passing through the hinge. An inclinometer provides measurements of such angles, and pressure sensors are located inside each muscle. This platform can be considered as a simplified version of a Stewart platform, a test bench on which control laws can be tried and evaluated before moving to more complex systems with more degrees of freedom.

\subsection{Model}

 This section reports  the differential equations describing the system dynamic and the different assumptions made.
A more detailed description of the complete model of the system is presented in \cite{saba2016complete}, with all the assumptions and explanations (including those concerning how the hysteresis has been taken into account).

\begin{table*}[t!] 
\begin{align} \label{eq:Mtxty}
\resizebox{0.9\hsize}{!}{$
M\left(\theta_x , \theta_y \right)=\displaystyle \scalemath{0.9}{\frac{R}{J} \begin{bmatrix}
-\sin\phi_1\cos\theta_x\cos\theta_y & -\sin\phi_2\cos\theta_x\cos\theta_y & -\sin\phi_3\cos\theta_x\cos\theta_y\\
\cos\phi_1\cos\theta_y+\sin\phi_1\sin\theta_x\sin\theta_y & \cos\phi_2\cos\theta_y+\sin\phi_2\sin\theta_x\sin\theta_y & \cos\phi_3\cos\theta_y+\sin\phi_3\sin\theta_x\sin\theta_y
\end{bmatrix}}$}
\end{align} \line(1,0){530} 
\begin{align} \label{eq:EG}
\left \{
\begin{array}{l}
\displaystyle\EMDD\\ \\
\displaystyle\GMDD\\
\end{array} \right.
\end{align} \line(1,0){530}
\end{table*}

The first elements to be modeled are the pneumatic muscles, which are supposed to be identical (they have the same length at rest \(l_0\), the same initial contraction \(\varepsilon_0\), etc.). 
The length contraction of each muscle (\(i=1,2,3\)) can be written as:
\begin{equation} \label{eq:eps}
\displaystyle\varepsilon_i=\frac{R}{l_{0}}\left(\cos\phi_i\sin\theta_y-\sin\phi_i\sin\theta_x\cos\theta_y\right) +\varepsilon_{0}
\end{equation}
Subsequently, the rate of contraction of each muscle is the time derivative of $\varepsilon_i$, i.e.
\begin{eqnarray}\label{eq:deps}
\begin{aligned}
\displaystyle\dot{\varepsilon}_i= & \frac{R}{l_{0}}\left[-\dot{\theta}_x\sin{\phi_i}\cos{\theta_x}\cos{\theta_y}\right.\\
& \left.+\dot{\theta}_y \left( \cos\phi_i\cos\theta_y+\sin\phi_i\sin\theta_x\sin\theta_y\right) \right]
\end{aligned}
\end{eqnarray}

\begin{equation} 
\begin{bmatrix}
\ddot{\theta}_x\\
\ddot{\theta}_y
\end{bmatrix}=M\left(\theta_x , \theta_y \right) \begin{bmatrix}
F_1\\
F_2\\
F_3
\end{bmatrix}+\frac{1}{J}\Gamma
\label{eq:MDD}
\end{equation}
where the matrix $M\left(\theta_x , \theta_y \right)$ is given in equation (\ref{eq:Mtxty}) at the top of the next page. The term \(\Gamma= [\Gamma_x, \Gamma_y]^\top\) contains the torques that will not be modelled (as an arbitrary choice) and will be left to the feedback control to take care of. Such terms are either due to friction, or to gyroscopic couplings between the two axes, or to external forces acting on the platform. The friction terms are quite difficult to model exactly, whereas the gyroscopic couplings are quite small due to the fact that the platform keeps always almost horizontal and moves at relatively low angular velocities. This allows writing the platform around each axis as decoupled, according to  \eqref{eq:MDD} above.
Such an equation can also be written as
\begin{equation}
\begin{bmatrix}
\ddot{\theta}_x\\
\ddot{\theta}_y
\end{bmatrix}=E\left(\theta_x,\theta_y\right)\begin{bmatrix} F_1 \\ F_2 \end{bmatrix} +G\left(\theta_x,\theta_y\right)F_3+\frac{1}{J}\Gamma
\end{equation}
where the matrices $E\left(\theta_x,\theta_y\right)$ and $G\left(\theta_x,\theta_y\right)$ are given in equation (\ref{eq:EG}) at the top of the next page. This form separates the effect of the first two forces with respect to \(F_3\), which makes it easier to approach the overactuation problem.

\begin{figure}[h]
\centering
\includegraphics[width=\columnwidth]{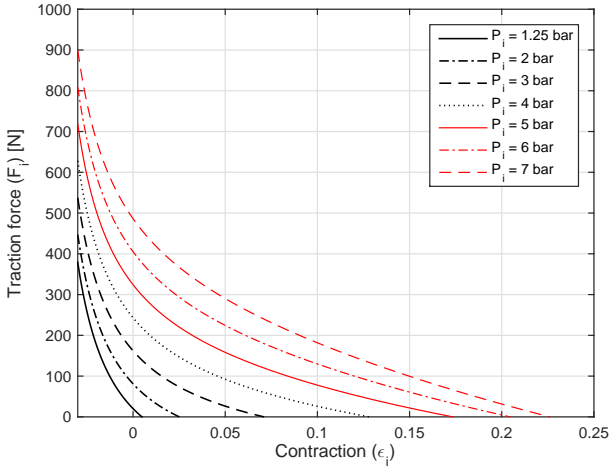} 
\caption{Traction force applied by a muscle as a function of the contraction $\varepsilon_i$ and absolute pressure $P_i$.} 
\label{fig:muscle}
\end{figure}

In turn, each force due to pneumatic muscles can be modeled with the so called quasi-static model \cite{daerden2002pneumatic,saba2016complete,fluidpower} as
\begin{equation} \label{eq:equasist}
F_i(P_i,\varepsilon_i)=H(\varepsilon_i)(P_i-P_0)+L(\varepsilon_i),
\end{equation}
where
\begin{equation} \label{eq: Leps}
L(\varepsilon_i)=K\frac{\varepsilon_i\left(\varepsilon_i-\varepsilon_a\right)}{\varepsilon_i+\varepsilon_b}
\end{equation}
and
\begin{equation} \label{eq: Heps}
H(\varepsilon_i)=\frac{\pi D_0^2}{4}\left[\frac{3\left(1-\varepsilon_i\right)^\alpha}{\tan^2\theta_0}-\frac{1}{\sin^2\theta_0}\right]
\end{equation} 
with $\alpha$, $K $, $\varepsilon_a$ and $\varepsilon_b$ experimentally determined constants. Considering that the operating range of the servovalves is for \(1.25~\mathrm{bar}\leqslant P_i \leqslant 7~\mathrm{bar}\), the possible forces for each muscle are represented in Figure~\ref{fig:muscle}. Notice that only traction forces are possible (the muscles cannot push).

The pressure inside each muscle is modeled as
\begin{equation} \label{eq:dynmcMuscle}
\dot{P}_i=\frac{krT}{V_i(\varepsilon_i)}\left[q_i(P_i,v_i)-\frac{P_i}{rT}\frac{\partial V(\varepsilon_i)}{\partial \varepsilon_i}\dot{\varepsilon}_i\right]
\end{equation}

where $k$ is the polytropic index of the gas, $r$ the perfect gas constant, $T$ the temperature (considered constant), $q_i$ the mass flow of gas, and $V_i$ the volume of the muscle, for which the following formula has been proposed \cite{saba2016complete,fluidpower}
\begin{equation}
\frac{\partial V}{\partial \varepsilon_i}\left(\varepsilon_i\right)=\frac{\pi}{4}D_0^2l_0\left[-\frac{1}{\sin^2\theta_0}+(\alpha+1)\frac{\left(1-\varepsilon_i\right)^\alpha}{\tan^2\theta_0}\right]
\end{equation}
where $D_0$, $l_0$ are the diameter and length of the muscle at rest, and $\theta_0$ is the weave angle of the muscle fibers (a constant).

\begin{figure}[h]
\centering
\includegraphics[width=\columnwidth]{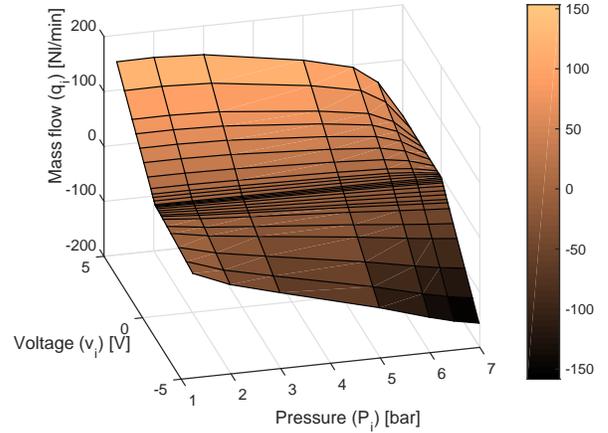} 
\caption{Mass flow of a servovalve as a function of voltage $v_i$ and absolute muscle pressure $P_i$.} 
\label{fig:servo}
\end{figure}

At last, the mass flow of gas $q_i$ entering each muscle is a nonlinear function of the pressure inside the muscle and the voltage $v_i$ fed to the servovalve. This function is considered as static, and it can be described by means of a polynomial approximation of experimental data \cite{olaby2005characterization} (graphically depicted in Figure~\ref{fig:servo}).

Considering that at each time instant, $P_i$ is measured by pressure sensors, it is possible to find the $v_i$ which gives the desired $q_i$ by a simple inversion of the polynomial function $q_i(P_i,v_i)$.


\subsection{Control objectives}

The aim of this testbed is to demonstrate the ability to track any smooth trajectory of \(\theta_x\) and \(\theta_y\). Trajectories of this kind can be chosen as  infinitely differentiable piece-wise polynomial function.
%

The angles of the platforms ($\theta_x$, $\theta_y$) are physically constrained to be in the range  $[-15^\circ, 15^\circ]$. For these values, for the contractions \(\varepsilon_i\) the ranges are constrained within \([-0.03, .21]\) (the muscles need to be contracted in order to apply a force), for which \(H(\varepsilon_i)\) is never equal to \(0\).

\section{Model analysis and control}
\label{sec:anal}
The accurate knowledge of the model allows the application of flatness-based control, at the condition of being able to prove that the system is flat. The relevant notions are recalled here.

\subsection{Flatness and flatness-based control}
The notion of flat system and flatness-based control for nonlinear systems have been introduced in  \cite{fliess1995flatness}. Basically, the ``flatness'' is a property of a dynamical system and a choice of its output \(y\), as defined here.

\begin{the:def}[Flat system - adapted from \cite{fliess1995flatness}]
A dynamical system of equations \(\dot{x} = f(x) + g(x) u\), with \(x \in \mathbb{R}^n\), \(u =\in \mathbb{R}^m\) is flat if there exist an \(\mathbb{R}^m\)-valued map \(h \), an \(\mathbb{R}^n\)-valued map  \(\eta\), and an \(\mathbb{R}^m\)-valued map \(\theta\) such that
\begin{equation}
y=h(x,u, u^{(1)},\ldots,  u^{(\nu)} )
\label{eq:flatness1}
\end{equation}
\begin{equation}
x=\eta(y, y^{(1)},\ldots,  y^{(\nu-1)} )
\label{eq:flatness2}
\end{equation}
\begin{equation}
u=\theta(y, y^{(1)},\ldots,  y^{(\nu)}  )
\label{eq:flatness3}
\end{equation}
for an appropriate value of  \(\nu \in \mathbb{N} \). The output \(y\) is then called ``flat output''.
\label{def:fla}
\end{the:def}

The idea of flatness can be explained briefly as follows. If one can choose as many output variables \(y_i\) as inputs (the system is square), such that it is possible to recover the state and the inputs from the derivatives of these output variables, then the system is flat and a flatness-based, open loop control law can be derived by system inversion (as explained later on). Two fundamental concepts for system inversion are characteristic index and coupling matrix.

\begin{the:def}[Characteristic index]
The characteristic index  of the \(i\)-th component \(y_i\) of \(y\) is the smallest \(\rho_i \in \mathbb{N}\) for which \(L_{g_j}L_f^{\rho_i-1} h_i \neq 0 \) for at least one value of \(j\).
\label{def:ci}
\end{the:def}

\begin{the:def}[Coupling matrix]
The coupling matrix \(\Delta(x)\) is given by the expression:
\begin{equation}
\Delta(x)\!\!=\!\!
\left[\!\!\!\!
\begin{array}{cccc}
L_{g_1}L_f^{\rho_1-1} h_1 & L_{g_2}L_f^{\rho_1-1} h_1 &\!\! \ldots \!\!& L_{g_m}L_f^{\rho_1-1} h_1 \\
L_{g_1}L_f^{\rho_2-1} h_2 & L_{g_2}L_f^{\rho_2-1} h_2 & \!\!\ldots & \!\!L_{g_m}L_f^{\rho_2-1} h_2 \\
\vdots & \vdots &\!\! \ddots\!\! & \vdots \\
L_{g_1}L_f^{\rho_m-1} h_m & L_{g_2}L_f^{\rho_m-1} h_m &\!\! \ldots & \!\!L_{g_m}L_f^{\rho_m-1} h_m 
\end{array}\!\!\!\!
\right].
\label{eq:cm}
\end{equation}
It can be shown that 
\begin{equation}
\left[\begin{array}{c} y_1^{(\rho_1)}\\ y_2^{(\rho_2)}\\\vdots \\ y_m^{(\rho_m)} \end{array}\right] =
\Delta(x) u + \left[\begin{array}{c} L^{\rho_1}_f h_1\\ L^{\rho_2}_f h_2 \\\vdots \\ L^{\rho_m}_f h_m \end{array}\right].
\label{eq:cmequation}
\end{equation}
\label{def:cm}
\end{the:def}

The control law which has been applied to the testbed is based on the following theorem, which is a well-known result for which no proof is necessary here.

\begin{the:the}[Adapted from \cite{fliess1995flatness}]
If a system of equations \(\dot{x}=f(x)+g(x)u\) with \(u \in \mathbb{R}^m\) is flat (Definition~\ref{def:fla}) with respect to a flat output \(y =h(x)\in \mathbb{R}^m\) with characteristic coefficients \(\rho_i\), and if the matrix \(\Delta(x)\) is invertible (at least locally), then it is possible to track a given smooth reference trajectory \({y}(t) =h({x}(t))\) by employing the control law
\begin{equation}
u=\Delta({x})^{-1}\left( \left[\begin{array}{c} {y}_1^{(\rho_1)}\\  {y}_2^{(\rho_2)}\\\vdots \\  {y}_m^{(\rho_m)} \end{array}\right]-
 \left[\begin{array}{c} L^{\rho_1}_f h_1\\ L^{\rho_2}_f h_2 \\\vdots \\ L^{\rho_m}_f h_m \end{array}\right]\right).
\label{eq:flatnesscontrol}
\end{equation}
\label{the:main}
\end{the:the}
%

It is possible to prove  that with this state trajectory, the system's dynamic of each \(y_i\) is linear  (simply a chain of \(\rho_i\) integrators).

\subsection{Complete state-space model}

The state of the platform model can be chosen as  $x=[x_1, x_2, x_3, \ldots x_7]^\top=[\theta_x , \theta_y ,\dot{\theta}_x , \dot{\theta}_y , P_1 , P_2, P_3]^\top$, whereas the input vector is  $u=[q_1 , q_2, q_3]^\top$. By neglecting the perturbation term $\Gamma$, the system dynamic can then be expressed as follows.
\begin{equation}
\dot{x}=f(x)+g(x)u
\end{equation}
where $f(x)=$
\begin{equation}
\scalemath{0.9}{
\begin{bmatrix}
x_3\\[10pt] x_4\\[10pt]
-\cos{x_1}\cos{x_2}\sin{\phi_1}\left(H\left(\varepsilon_1\right)(x_5-P_0)+L\left(\varepsilon_1\right)\right)\\
-\cos{x_1}\cos{x_2}\sin{\phi_2}\left(H\left(\varepsilon_2\right)(x_6-P_0)+L\left(\varepsilon_2\right)\right)\\
-\cos{x_1}\cos{x_2}\sin{\phi_3}\left(H\left(\varepsilon_3\right)(x_7-P_0)+L\left(\varepsilon_3\right)\right)
\\
\\[10pt]
\quad\left(\cos{\phi_1}\cos{x_2}+\sin{\phi_1}\sin{x_1}\sin{x_2}\right)\left(H\left(\varepsilon_1\right)(x_5-P_0)+L\left(\varepsilon_1\right)\right)\\
+\left(\cos{\phi_2}\cos{x_2}+\sin{\phi_2}\sin{x_1}\sin{x_2}\right)\left(H\left(\varepsilon_2\right)(x_6-P_0)+L\left(\varepsilon_2\right)\right)\\
+\left(\cos{\phi_3}\cos{x_2}+\sin{\phi_3}\sin{x_1}\sin{x_2}\right)\left(H\left(\varepsilon_3\right)(x_7-P_0)+L\left(\varepsilon_3\right)\right)\\
\\[10pt]
a(\varepsilon_1,\dot{\varepsilon}_1)(x_5-P_0)\\[10pt]
a(\varepsilon_2,\dot{\varepsilon}_2)(x_6-P_0)\\[10pt]
a(\varepsilon_3,\dot{\varepsilon}_3)(x_7-P_0)
\end{bmatrix}},
\end{equation}
$g(x)=[g_1(x), g_2(x), g_3(x)]$ with
\begin{equation}
g_1(x)=\begin{bmatrix}
0\\0\\0\\0\\b(\varepsilon_1)\\0\\0
\end{bmatrix}, \quad g_2(x)=\begin{bmatrix}
0\\0\\0\\0\\0\\b(\varepsilon_2)\\0
\end{bmatrix}, \quad g_3(x)=\begin{bmatrix}
0\\0\\0\\0\\0\\0\\b(\varepsilon_3)
\end{bmatrix}
\end{equation}
with
\begin{equation}
\left \{
\begin{array}{l}
\displaystyle a\left(\varepsilon_i,\dot{\varepsilon}_i\right)=-\frac{k}{V(\varepsilon_i)}\frac{\partial V(\varepsilon_i)}{\partial \varepsilon_i}\dot{\varepsilon}_i\\
\displaystyle b(\varepsilon_i)=\frac{krT}{V(\varepsilon_i)}
\end{array}
\right.
\end{equation}

\subsection{Flatness of the model}

The system is flat if  a vector flat output \([y_1\, y_2\, y_3]^\top\), according to \eqref{eq:flatness1}, can be found. Such flat output has to fulfill both \eqref{eq:flatness2}, i.e., it should be possible to express the state vector as a function of its time derivatives, and  \eqref{eq:flatness3}, i.e. it should be possible to express the input as a function of its derivatives.

This paragraph  shows that the choice
\begin{equation}
\left \{ \begin{array}{l}
y_1=x_1\\
y_2=x_2\\
y_3=F_3=H(\varepsilon_3)(x_7-P_0)+L(\varepsilon_3)\\
\end{array}
\right.
\end{equation}
actually works in making the system flat.

First consider condition \eqref{eq:flatness2}; $x_1$, $x_2$, $x_3$ and $x_4$ can be obtained directly from $y_1$, $y_2$ and their first degree time derivatives. Once $x_1$, $x_2$, $x_3$ and $x_4$ are known, all $\varepsilon_i$, $H(\varepsilon_i)$ and $L(\varepsilon_i)$ are determined as well. Since \(F_3\) is an output and \(H(\varepsilon_3)\neq 0\), $x_7$ is immediately also determined. At last, \(x_5\) and \(x_6\) can be determined from \(\dot{x}_3\) and \(\dot{x}_4\) if the matrix
$$
\scalemath{0.85}{\left[\!
\begin{array}{cc}
-\cos x_1 \cos x_2  \sin \phi_1  &  -\cos x_1 \cos x_2  \sin \phi_2 \\ 
\cos x_2 \cos \phi_1\! +\! \sin x_1 \sin x_2 \sin \phi_1 \! &\! \cos x_2 \cos \phi_2 \!+\! \sin x_1 \sin x_2 \sin \phi_2  \\
\end{array}
\!\right]}
$$
is invertible. The determinant of this matrix is $\cos^2 x_2 \cos x_1(\sin \phi_2 \cos \phi_1 -\sin \phi_1 \cos \phi_2)$
which is never \(0\) in the range of \(\theta_x=x_1\), \(\theta_y=x_2\) allowed for the platform (i.e. they never reach \(\pm 90^\circ\)).

Secondarily, consider condition \eqref{eq:flatness2}; a necessary condition for this is that the sum of the characteristic indices of the three outputs is the same as the number of states, i.e. \(7\). The computation of such indices leads to the following results.

\phantom{a}

\begin{itemize}

\item{Output $y_1$}
\vspace{.3cm}

$
L_{g_1}y_1= L_{g_2}y_1= L_{g_3}y_1=0 \Rightarrow \rho_1>1$;\vspace{.3cm}

$L_{f}y_1=x_3$;

$L_{g_1}L_{f}y_1= L_{g_2}L_{f}y_1= L_{g_3}L_{f}y_1=0 \quad \Rightarrow \rho_1>2$;\vspace{.3cm}

$L_f^2y_1=\dot{x}_3=  
-\cos{x_1}\cos{x_2}(\sin{\phi_1}(H(\varepsilon_1)(x_5-P_0)+
$

$
L(\varepsilon_1))
+\sin{\phi_2}(H(\varepsilon_2)(x_6-P_0)\!+\!L(\varepsilon_2))+$ $
\sin{\phi_3}(H(\varepsilon_3)(x_7-P_0)\!+\!L(\varepsilon_3)));
$

$
L_{g_1}L_f^2y_1=-\sin{\phi_1}\cos{x_1}\cos{x_2}H(\varepsilon_1)b(\varepsilon_1)\\
L_{g_2}L_f^2y_1=-\sin{\phi_2}\cos{x_1}\cos{x_2}H(\varepsilon_2)b(\varepsilon_2)\\
L_{g_3}L_f^2y_1=-\sin{\phi_3}\cos{x_1}\cos{x_2}H(\varepsilon_3)b(\varepsilon_3)
$ \vspace{.3cm}

It can be pointed out that $L_{g_1}L_f^2y_1$, $L_{g_2}L_f^2y_1$  and  $L_{g_3}L_f^2y_1$ are never equal to $0$ for the $x_1$ and $x_2$ within the valid range, so $\rho_1=3$.\vspace{.3cm}

\item{Output $y_2$}\vspace{.3cm}

$
L_{g_1}y_2= L_{g_2}y_2= L_{g_3}y_2=0 \Rightarrow \rho_2>1;$\vspace{.3cm}

$
L_{f}y_2=x_4$; \vspace{.3cm}

$
L_{g_1}L_{f}y_2= L_{g_2}L_{f}y_2= L_{g_3}L_{f}y_2=0 \Rightarrow \rho_2>2$;  \vspace{.3cm}

$
L_f^2y_2 = \dot{x}_4=$

$ (\cos{\phi_1}\cos{x_2}+\sin{\phi_1}\sin{x_1}\sin{x_2})(H(\varepsilon_1)(x_5-P_0)+L(\varepsilon_1))
+(\cos{\phi_2}\cos{x_2}+\sin{\phi_2}\sin{x_1}\sin{x_2})(H(\varepsilon_2)(x_6-P_0)+L(\varepsilon_2))
+(\cos{\phi_3}\cos{x_2}+\sin{\phi_3}\sin{x_1}\sin{x_2})(H(\varepsilon_3)(x_7-P_0)+L(\varepsilon_3));
$

$
L_{g_1}L_f^2y_2=
$

 $(\cos{\phi_1}\cos{x_2}+\sin{\phi_1}\sin{x_1}\sin{x_2})H(\varepsilon_1)b(\varepsilon_1)$
 
 $
L_{g_2}L_f^2y_2=$

$(\cos{\phi_2}\cos{x_2}+\sin{\phi_2}\sin{x_1}\sin{x_2})H(\varepsilon_2)b(\varepsilon_2)$

$
L_{g_3}L_f^2y_2=$

 $(\cos{\phi_3}\cos{x_2}+\sin{\phi_3}\sin{x_1}\sin{x_2})H(\varepsilon_3)b(\varepsilon_3)
$ \vspace{.3cm}

Notice that $L_{g_2}L_f^2y_2$ can never be zero in the valid range, as the function
\begin{equation}
z=\cos{\phi_2}\cos{x_2}+\sin{\phi_2}\sin{x_1}\sin{x_2}
\end{equation}
plotted in Figure \ref{fig:Lg2} never reaches zero in this interval. So $\rho_2=3$.\vspace{.3cm}

\begin{figure}[h]
\includegraphics[width=\columnwidth]{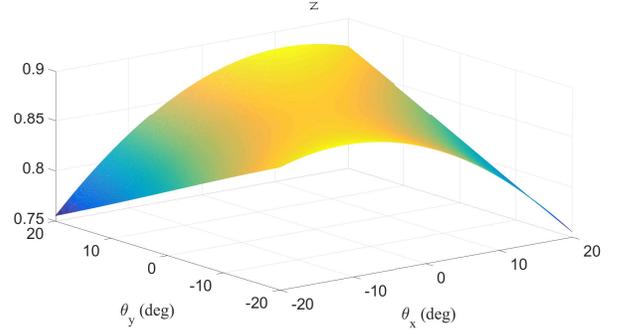} 
\caption{Value of $z$ as function of $x_1=\theta_x$ and $x_2=\theta_y$.} 
\label{fig:Lg2}
\end{figure}

\item{Output $y_3$}\vspace{.3cm}

$
L_{g_1}y_3= L_{g_2}y_3=0$;

$L_{g_3}y_3=
b(\varepsilon_3)H(\varepsilon_3)\neq0 
\Rightarrow \rho_3=1.
$\\

\end{itemize}

The necessary condition of \(\rho_1+\rho_2+\rho_3=7\) is satisfied. The last step is to verify that the decoupling matrix 
\begin{equation}
\Delta=\begin{bmatrix}
L_{g_1}L_f^2y_1 & L_{g_2}L_f^2y_1 & L_{g_3}L_f^2y_1\\
L_{g_1}L_f^2y_2 & L_{g_2}L_f^2y_2 & L_{g_3}L_f^2y_2\\
L_{g_1}y_3 & L_{g_2}y_3 & L_{g_3}y_3\\ 
\end{bmatrix}
\end{equation}
is invertible. 
The expression of $\Delta$ is made explicit in (\ref{eq:Delta}) at the top of the next page (with the shorthand notation of $H_i = H(\varepsilon_i)$, $L_i = L(\varepsilon_i)$, $b_i = b(\varepsilon_i)$).
\begin{table*}[t!] 
\begin{align} \label{eq:Delta}
\Delta=\begin{bmatrix}
-\sin{\phi_1}\cos{x_1}\cos{x_2}H_1 b_1 & -\sin{\phi_2}\cos{x_1}\cos{x_2}H_2 b_2\
& -\sin{\phi_3}\cos{x_1}\cos{x_2}H_3 b_3\\
 \left(\cos{\phi_1}\cos{x_2}+\sin{\phi_1}\sin{x_1}\sin{x_2}\right)H_1 b_1 & \left(\cos{\phi_2}\cos{x_2}+\sin{\phi_2}\sin{x_1}\sin{x_2}\right)H_2b_2 & \left(\cos{\phi_3}\cos{x_2}+\sin{\phi_3}\sin{x_1}\sin{x_2}\right)H_3b_3\\
0 & 0 & H_3b_3\\
\end{bmatrix}
\end{align}\line(1,0){530}
\end{table*}

\begin{figure}
\includegraphics[width=\columnwidth]{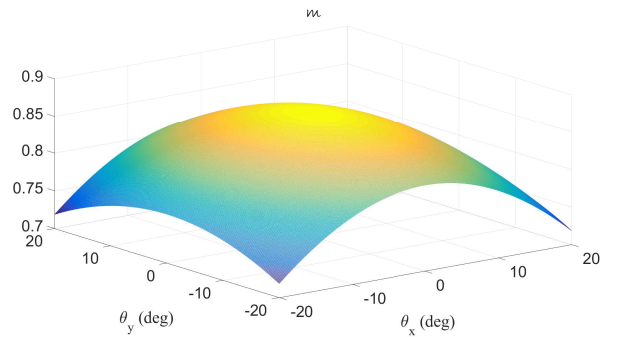} 
\caption{Values of $m$ as function of $\theta_x$ and $\theta_y$.} 
\label{fig:delta}
\end{figure}

The determinant of this matrix is
$|\Delta|=  H_1 H_2 H_3 b_1 b_2 b_3 m
$
with 
%
$m=  -\sin{\phi_1}\cos{x_1}\cos{x_2}(\cos{\phi_2}\cos{x_2}+\sin{\phi_2}\sin{x_1}\sin{x_2})  +\sin{\phi_2}\cos{x_1}\cos{x_2}(\cos{\phi_1}\cos{x_2}+\sin{\phi_1}\sin{x_1}\sin{x_2}).
$
The values of $m$ as function of $\theta_x$ and $\theta_y$ in the valid interval are depicted in Figure \ref{fig:delta}

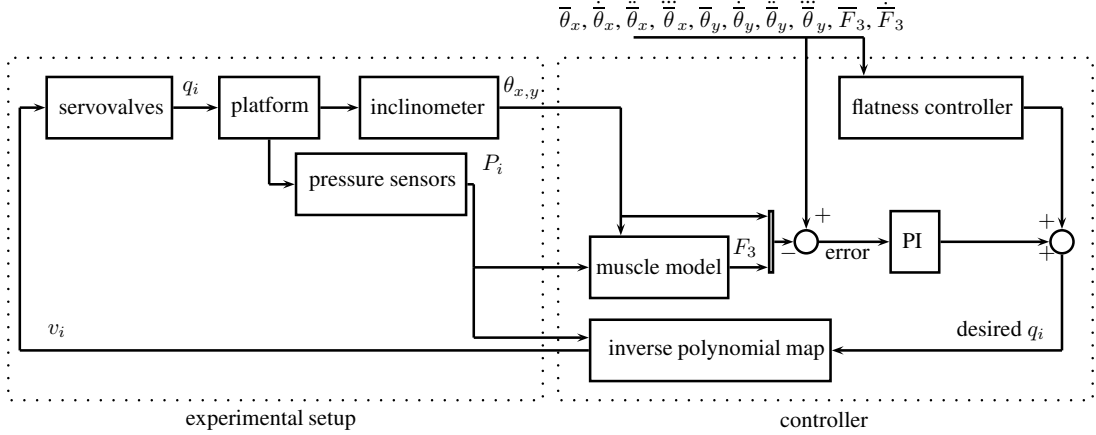
\begin{figure*}[t!]
\centering
\scalebox{.85} 
{
\begin{pspicture}(0,-3.3576562)(26.522812,3.3576562)
\psframe[linewidth=0.04,dimen=outer](6.7809377,2.2592187)(5.1809373,1.2592187)
\usefont{T1}{ptm}{m}{n}
\rput(5.9907813,1.7692188){platform}
\psframe[linewidth=0.04,dimen=outer](9.580937,2.2592187)(7.3809376,1.2592187)
\usefont{T1}{ptm}{m}{n}
\rput(8.456718,1.7692188){inclinometer}
\psframe[linewidth=0.04,dimen=outer](17.780937,2.2592187)(14.880938,1.2592187)
\usefont{T1}{ptm}{m}{n}
\rput(16.345156,1.7692188){flatness controller}
\psframe[linewidth=0.04,dimen=outer](9.080937,1.0592188)(6.3809376,0.05921875)
\usefont{T1}{ptm}{m}{n}
\rput(7.7553124,0.56921875){pressure sensors}
\psline[linewidth=0.04cm,arrowsize=0.05291667cm 2.0,arrowlength=1.4,arrowinset=0.4]{->}(6.7809377,1.7592187)(7.3809376,1.7592187)
\psline[linewidth=0.04cm](5.9809375,0.55921876)(5.9809375,1.2592187)
\psline[linewidth=0.04cm](11.680938,2.8592188)(15.280937,2.8592188)
\psframe[linewidth=0.04,dimen=outer](13.180938,-0.24078125)(10.980938,-1.2407813)
\usefont{T1}{ptm}{m}{n}
\rput(12.075625,-0.73078126){muscle model}
\psline[linewidth=0.04cm,arrowsize=0.05291667cm 2.0,arrowlength=1.4,arrowinset=0.4]{->}(5.9809375,0.55921876)(6.3809376,0.55921876)
\psline[linewidth=0.04cm](9.180938,0.55921876)(9.180938,-0.74078125)
\psline[linewidth=0.04cm,arrowsize=0.05291667cm 2.0,arrowlength=1.4,arrowinset=0.4]{->}(9.180938,-0.74078125)(10.980938,-0.74078125)
\psline[linewidth=0.04cm,arrowsize=0.05291667cm 2.0,arrowlength=1.4,arrowinset=0.4]{->}(11.480938,1.7592187)(11.480938,-0.24078125)
\psline[linewidth=0.04cm,arrowsize=0.05291667cm 2.0,arrowlength=1.4,arrowinset=0.4]{->}(13.180938,-0.74078125)(13.780937,-0.74078125)
\psline[linewidth=0.04cm](9.080937,0.55921876)(9.180938,0.55921876)
\psline[linewidth=0.04cm,arrowsize=0.05291667cm 2.0,arrowlength=1.4,arrowinset=0.4]{->}(11.480938,0.05921875)(13.780937,0.05921875)
\psframe[linewidth=0.04,dimen=outer](13.880938,0.15921874)(13.780937,-0.8407813)
\psline[linewidth=0.04cm,arrowsize=0.05291667cm 2.0,arrowlength=1.4,arrowinset=0.4]{->}(15.280937,2.8592188)(15.280937,2.2592187)
\psline[linewidth=0.04cm,arrowsize=0.05291667cm 2.0,arrowlength=1.4,arrowinset=0.4]{->}(14.380938,2.8592188)(14.380938,-0.14078125)
\pscircle[linewidth=0.04,dimen=outer](14.380938,-0.34078124){0.2}
\psline[linewidth=0.04cm,arrowsize=0.05291667cm 2.0,arrowlength=1.4,arrowinset=0.4]{->}(13.880938,-0.34078124)(14.180938,-0.34078124)
\usefont{T1}{ptm}{m}{n}
\rput(14.6423435,0.06921875){$+$}
\usefont{T1}{ptm}{m}{n}
\rput(14.102344,-0.53078127){$-$}
\psline[linewidth=0.04cm,arrowsize=0.05291667cm 2.0,arrowlength=1.4,arrowinset=0.4]{->}(14.580937,-0.34078124)(15.680938,-0.34078124)
\psline[linewidth=0.04cm,arrowsize=0.05291667cm 2.0,arrowlength=1.4,arrowinset=0.4]{->}(16.480938,-0.34078124)(18.180937,-0.34078124)
\pscircle[linewidth=0.04,dimen=outer](18.380938,-0.34078124){0.2}
\usefont{T1}{ptm}{m}{n}
\rput(18.142344,-0.03078125){$+$}
\usefont{T1}{ptm}{m}{n}
\rput(18.142344,-0.53078127){$+$}
\psline[linewidth=0.04cm](17.780937,1.7592187)(18.380938,1.7592187)
\psline[linewidth=0.04cm,arrowsize=0.05291667cm 2.0,arrowlength=1.4,arrowinset=0.4]{<-}(18.380938,-0.14078125)(18.380938,1.7592187)
\psframe[linewidth=0.04,dimen=outer](16.480938,0.15921874)(15.680938,-0.8407813)
\usefont{T1}{ptm}{m}{n}
\rput(16.043594,-0.33078125){PI}
\psline[linewidth=0.04cm](18.380938,-0.54078126)(18.380938,-2.0407813)
\psline[linewidth=0.04cm,arrowsize=0.05291667cm 2.0,arrowlength=1.4,arrowinset=0.4]{->}(18.380938,-2.0407813)(14.780937,-2.0407813)
\psline[linewidth=0.04cm](10.980938,-2.0407813)(2.0809374,-2.0407813)
\psline[linewidth=0.04cm](2.0809374,1.7592187)(2.0809374,-2.0407813)
\psline[linewidth=0.04cm,arrowsize=0.05291667cm 2.0,arrowlength=1.4,arrowinset=0.4]{->}(2.0809374,1.7592187)(2.4809375,1.7592187)
\psline[linewidth=0.04cm,arrowsize=0.05291667cm 2.0,arrowlength=1.4,arrowinset=0.4]{->}(4.4809375,1.7592187)(5.1809373,1.7592187)
\psline[linewidth=0.04cm,arrowsize=0.05291667cm 2.0,arrowlength=1.4,arrowinset=0.4]{->}(9.180938,-1.8407812)(10.980938,-1.8407812)
\psframe[linewidth=0.04,dimen=outer](14.780937,-1.5407813)(10.980938,-2.5407813)
\usefont{T1}{ptm}{m}{n}
\rput(12.979531,-2.0307813){inverse polynomial map}
\psframe[linewidth=0.04,dimen=outer](4.4809375,2.2592187)(2.4809375,1.2592187)
\usefont{T1}{ptm}{m}{n}
\rput(3.5059376,1.7692188){servovalves}
\usefont{T1}{ptm}{m}{n}
\rput(2.6623437,-1.7307812){$v_i$}
\usefont{T1}{ptm}{m}{n}
\rput(4.762344,2.0692186){$q_i$}
\usefont{T1}{ptm}{m}{n}
\rput(9.9323435,2.0692186){$\theta_{x,y}$}
\psline[linewidth=0.04cm](9.180938,-0.74078125)(9.180938,-0.74078125)
\usefont{T1}{ptm}{m}{n}
\rput(9.482344,0.86921877){$P_i$}
\usefont{T1}{ptm}{m}{n}
\rput(13.422344,-0.43078125){$F_3$}
\usefont{T1}{ptm}{m}{n}
\rput(13.222343,3.1692188){$\overline{\theta}_x$, $\dot{\overline{\theta}}_x$, $\ddot{\overline{\theta}}_x$, $\dddot{\overline{\theta}}_x$,  $\overline{\theta}_y$, $\dot{\overline{\theta}}_y$, $\ddot{\overline{\theta}}_y$,  $\dddot{\overline{\theta}}_y$, $\overline{F}_3$,  $\dot{\overline{F}}_3$}
\usefont{T1}{ptm}{m}{n}
\rput(15.025312,-0.53078127){error}
\usefont{T1}{ptm}{m}{n}
\rput(17.425938,-1.7307812){desired $q_i$}
\psline[linewidth=0.04cm](9.580937,1.7592187)(11.480938,1.7592187)
\psline[linewidth=0.04cm](9.180938,-0.74078125)(9.180938,-1.8407812)
\psframe[linewidth=0.04,linestyle=dotted,dotsep=0.16cm,dimen=outer](18.880938,2.5592186)(10.480938,-2.8407812)
\psframe[linewidth=0.04,linestyle=dotted,dotsep=0.16cm,dimen=outer](10.280937,2.5592186)(1.8809375,-2.8407812)
\usefont{T1}{ptm}{m}{n}
\rput(6.008125,-3.1307812){experimental setup}
\usefont{T1}{ptm}{m}{n}
\rput(14.655937,-3.1307812){controller}
\end{pspicture} 
}\caption{Global control scheme.}
\label{fig:scheme}\line(1,0){530}
\end{figure*}

Accordingly, $|\Delta|\neq 0$, so the decoupling matrix is invertible over the operating range and the chosen output is proven to be flat.

\subsection{Control law}

The platform can then be controlled in open-loop with the law
\begin{equation}
q=\Delta^{-1}(\overline{x})\left(
\left[\begin{array}{c} \overline{y}_1^{(\rho_1)}\\  \overline{y}_2^{(\rho_2)}\\  \overline{y}_3^{(\rho_3)} \end{array}\right]
- \begin{bmatrix}
L_f^3\overline{y}_1\\L_f^3\overline{y}_2\\L_f\overline{y}_3
\end{bmatrix}\right)+\Delta^{-1}(\overline{x})w
\end{equation}
where $\overline{y}_i$ is the desired trajectory and $w$ is an additional control term. Due to the presence of the perturbation terms which have been neglected ($\Gamma$), if $w=0$ there will necessarily be a non-zero error $\epsilon_i=y_i-\overline{y}_i$. Considering that the dynamic of the system under this law is just a chain of integrators, by imposing to $w$ a feedback law as a function of the $\epsilon_i$ closes the loop. In this case, a proportional-integral controller (PI) has been tuned. Figure~\ref{fig:scheme} shows the overall control scheme.

The flatness-based control is for some aspects, quite similar to feedback linearisation control \cite{isidori2013nonlinear}. The main differences lie in the fact that flatness-based requires a specific choice of flat output (whereas feedback linearisation can take any output, assuming the system is observable), and that it does not require a knowledge or measure of the state variable. On the other hand, the baseline flatness-based control is feedforward only, which requires the introduction of the additional feedback term $w$.

\section{Solving the overactuation}
\label{sec:over}
It can be pointed out that the platform is overactuated, in the sense that the three forces applied by the muscles are generating only two torques. To tell it in another way, the average value of the $F_i$ is irrelevant for the platform's dynamic; if a given $F_1=\tilde{F}_1$, $F_2=\tilde{F}_2$ and $F_3=\tilde{F}_3$ generate certain torques, then  $F_1=\tilde{F}_1+F_0$, $F_2=\tilde{F}_2+F_0$ and $F_3=\tilde{F}_3+F_0$ will generate the same torques for any $F_0$. On the other hand, it is useful to have three muscles instead of two due to the fact that muscles can only pull and not push, i.e. their force range is quite limited as shown by Figure~\ref{fig:muscle}.

\begin{figure}[h]
\centering
\includegraphics[width=\columnwidth]{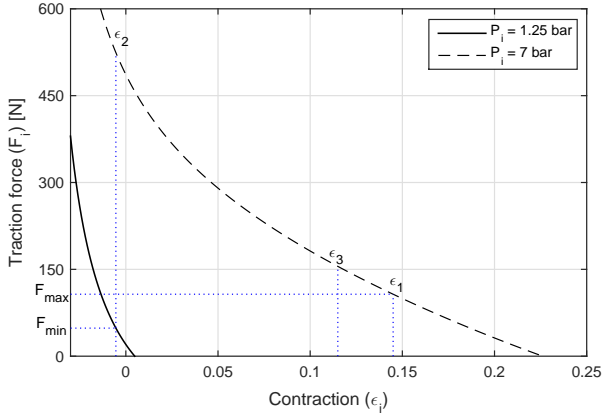} 
\caption{For a given position, the three contractions $\varepsilon_i$ are given, so not all forces are possible for each muscle, but only those within the pressure range between $1.25$ and $7$~bar. One can compute $F_{max}$ (the maximum force that all muscles can exert) and $F_{max}$ (the minimum force that all muscles can exert). Setting $F_3$ as the average between these two allows all muscles to apply the desired forces and to maximise their range.} 
\label{fig:over}
\end{figure}

The choice of $F_3$ as one of the flat output can be then interpreted in the light of this fact: first of all, it would be impossible to resolve the state from the output if one of the forces (or pressure) is not measured, as their effects on the angles is the same up to a constant term. Secondarily, the flat control allows choosing the value of $F_3$, which lets one choose the best value in order to let all the muscles be in their valid force range. Consider that at each instant, the $\varepsilon_i$ are determined by the instantaneous geometry, which implies that each muscle has a limited interval of possible applicable forces (see Figure~\ref{fig:over}). One can find the intersection of such intervals and call $F_{min}$ its minimum and $F_{max}$ its maximum. Under the reasonable hypothesis that the platform turns slowly (in any case it is constrained to angles smaller than $15^\circ$), it can be assumed that $F_1$, $F_2$ and $F_3$ have to be close to the equilibrium values, i.e. $F_1 \approx F_2 \approx F_3$. For this reason, setting the reference for $F_3$ as
\begin{equation}
\label{eq:F3}
\overline{F}_3=\frac{1}{2}(F_{max}(\varepsilon_1,\varepsilon_2,\varepsilon_3)+F_{min}(\varepsilon_1,\varepsilon_2,\varepsilon_3))
\end{equation}
gives the best chances of having $F_2$ and $F_3$ within the realisable interval as well.

For the experiment in the next section, the reference for $F_3$ has been determined by the law in \eqref{eq:F3}.

\section{Experimental results}
\label{sec:expe}

\begin{figure}[h]
\centering
\includegraphics[width=\columnwidth]{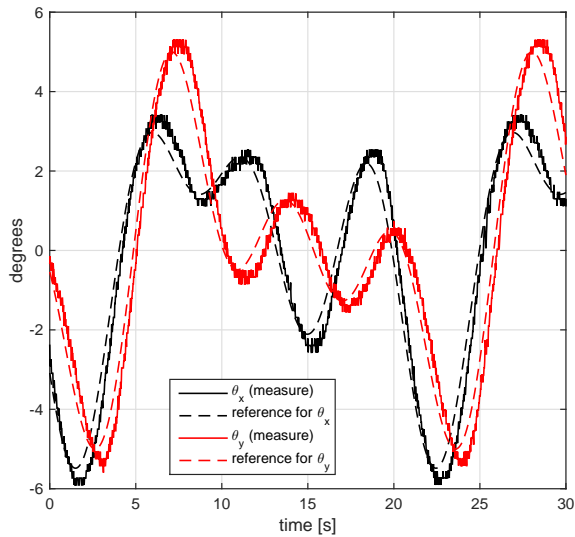} 
\caption{Trajectory tracking with simple PI control. } 
\label{fig:pi}
\end{figure}

\begin{figure}[h]
\centering
\includegraphics[width=\columnwidth]{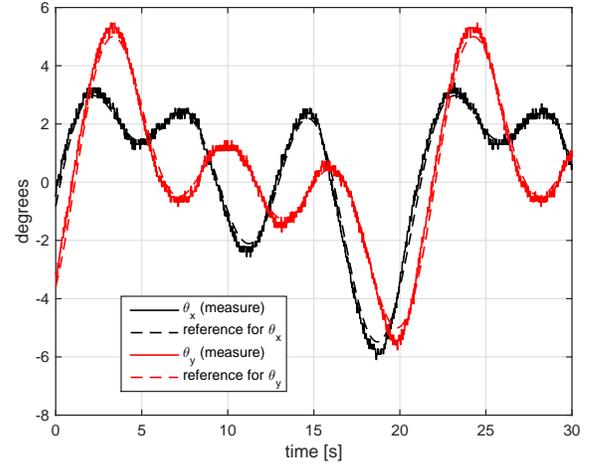} 
\caption{Trajectory tracking with flatness-based control plus a PI. } 
\label{fig:flat1}
\end{figure}

In order to assess the performance of the proposed control approach, an experiment has been conducted on the platform. The proposed flatness-based control coupled with PI has been compared to a PI controller, empirically tuned to get the best apparent performances. The same reference trajectory (a combination of sinusoids) has been tested for both controllers. Figure~\ref{fig:pi} reports the results for the PI controller, whereas Figure~\ref{fig:flat1} shows the flatness-based controller results. It is apparent from the picture that the feedforward action added by the flatness-based control greatly improves the tracking ability; in fact, it can be computed that the root mean square tracking errors (for $\theta_x$ and $\theta_y$ respectively) are $0.51$ and $0.59$~degrees for the PI case. With the flatness controller, these root mean square errors become less than half, i.e. $0.25$ and $0.29$~degrees respectively (consider also that the inclinometers' output has a quantisation equivalent to $0.18$~degrees).

\begin{figure}[h]
\centering
\includegraphics[width=\columnwidth]{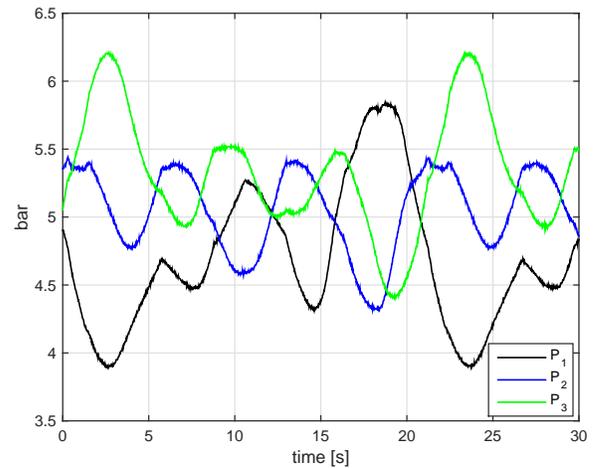} 
\caption{Pressures inside the pneumatic artificial muscles during the flatness-based control plus PI experiment.} 
\label{fig:flatpressures}
\end{figure}

\begin{figure}[h]
\centering
\includegraphics[width=\columnwidth]{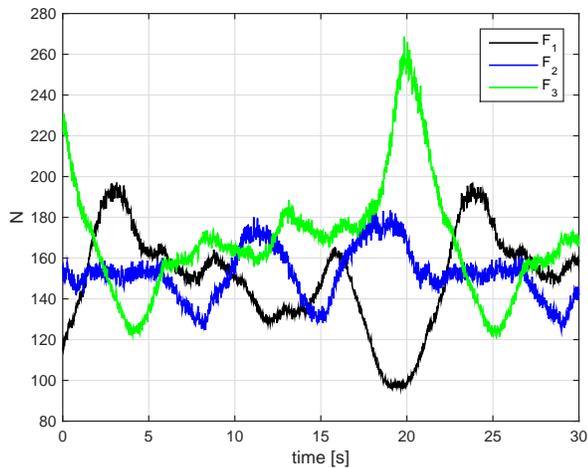} 
\caption{Forces applied by the pneumatic artificial muscles (estimated from the pressures and the contractions) with flatness-based control plus a PI.} 
\label{fig:flatforce}
\end{figure}

Figure~\ref{fig:flatpressures} shows the evolution of the pressures during the flatness-based controller test. Figure~\ref{fig:flatforce} shows the force of the three muscles during the same test; remember that the reference for $F_3$ is determined with the overactuation-solving law proposed in Section~\ref{sec:over}. Notice that the forces never saturate (and neither do the voltages or the pressure), which validates the proposed strategy.


%

\section{Conclusion}
\label{sec:conc}
This paper has presented the successful application of a flatness-based controller to a  platform featuring three PAMs. The experimental results clearly show a better trajectory tracking compared to a simple PI controller.
Future research will look at the possibility of using PAMs for building a complete six-degree-of-freedom Stewart platform, and controlling it with the same approach.

\bibliographystyle{plain}
\bibliography{david2017}

\begin{thebibliography}{10}

\bibitem{aschemann2008sliding}
H.~Aschemann and D.~Schindele.
\newblock Sliding-mode control of a high-speed linear axis driven by pneumatic
  muscle actuators.
\newblock {\em IEEE Transactions on Industrial Electronics}, 55(11):3855--3864,
  2008.

\bibitem{ba2016integrated}
D.X. Ba, T.Q. Dinh, and K.K. Ahn.
\newblock An integrated intelligent nonlinear control method for a pneumatic
  artificial muscle.
\newblock {\em IEEE/ASME Transactions on Mechatronics}, 21(4):1835--1845, 2016.

\bibitem{fluidpower}
E.~Bideaux, S.~Sermeno~Mena, and S.~Sesmat.
\newblock Parallel manipulator driven by pneumatic muscles.
\newblock In {\em 8th International Conference on Fluid Power (8th IFK)},
  Dresden, Germany, 2012.

\bibitem{saba2016complete}
D.~Bou~Saba, E.~Bideaux, X.~Brun, and P.~Massioni.
\newblock A complete model of a two degree of freedom platform actuated by
  three pneumatic muscles elaborated for control synthesis.
\newblock In {\em BATH/ASME 2016 Symposium on Fluid Power and Motion Control},
  pages V001T01A004--V001T01A004. American Society of Mechanical Engineers,
  2016.

\bibitem{cai2000sliding}
D.~Cai and Y.~Dai.
\newblock A sliding mode controller for manipulator driven by artificial muscle
  actuator.
\newblock In {\em Proceedings of the 2000 IEEE International Conference on
  Control Applications}, pages 668--673. IEEE, 2000.

\bibitem{chou1996measurement}
C.-P. Chou and B.~Hannaford.
\newblock Measurement and modeling of {McKibben} pneumatic artificial muscles.
\newblock {\em IEEE Transactions on Robotics and Automation}, 12(1):90--102,
  1996.

\bibitem{daerden2002pneumatic}
F.~Daerden and D.~Lefeber.
\newblock Pneumatic artificial muscles: actuators for robotics and automation.
\newblock {\em European journal of mechanical and environmental engineering},
  47(1):11--21, 2002.

\bibitem{fliess1995flatness}
M.~Fliess, J.~L{\'e}vine, P.~Martin, and P.~Rouchon.
\newblock Flatness and defect of non-linear systems: introductory theory and
  examples.
\newblock {\em International Journal of Control}, 61(6):1327--1361, 1995.

\bibitem{isidori2013nonlinear}
A.~Isidori.
\newblock {\em Nonlinear control systems}.
\newblock Springer Science \& Business Media, 2013.

\bibitem{olaby2005characterization}
O.~Olaby, X.~Brun, S.~Sesmat, T.~Redarce, and E.~Bideaux.
\newblock Characterization and modeling of a proportional value for control
  synthesis.
\newblock In {\em Proceedings of the JFPS International Symposium on Fluid
  Power}, volume 2005, pages 771--776. The Japan Fluid Power System Society,
  2005.

\bibitem{abd2016design}
R.A. Rahman and N.~Sepehri.
\newblock Design and experimental evaluation of a dynamical adaptive
  backstepping-sliding mode control scheme for positioning of an
  antagonistically paired pneumatic artificial muscles driven actuating system.
\newblock {\em International Journal of Control}, pages 1--26, 2016.

\bibitem{robinson2016nonlinear}
R.M. Robinson, C.S. Kothera, R.M. Sanner, and N.M. Wereley.
\newblock Nonlinear control of robotic manipulators driven by pneumatic
  artificial muscles.
\newblock {\em IEEE/ASME Transactions on Mechatronics}, 21(1):55--68, February
  2016.

\bibitem{shen2010nonlinear}
X.~Shen.
\newblock Nonlinear model-based control of pneumatic artificial muscle servo
  systems.
\newblock {\em Control Engineering Practice}, 18(3):311--317, 2010.

\bibitem{shi2013hybrid}
G.L. Shi and W.~Shen.
\newblock Hybrid control of a parallel platform based on pneumatic artificial
  muscles combining sliding mode controller and adaptive fuzzy {CMAC}.
\newblock {\em Control Engineering Practice}, 1(1):76--86, 2013.

\bibitem{tondu2000modeling}
B.~Tondu and P.~Lopez.
\newblock Modeling and control of {McKibben} artificial muscle robot actuators.
\newblock {\em Control Systems, IEEE}, 20(2):15--38, 2000.

\bibitem{zhu2008adaptive}
X.~Zhu, G.~Tao, B.~Yao, and J.~Cao.
\newblock Adaptive robust posture control of parallel manipulator driven by
  pneumatic muscles with redundancy.
\newblock {\em IEEE/ASME Transactions on Mechatronics}, 13(4):441--450, August
  2008.

\end{thebibliography}

\end{document}